# Surface Deformation During an Action Potential in Pearled Cells

Matan Mussel[a,b], Christian Fillafer[a], Gal Ben-Porath[c], Matthias F. Schneider[a,*]

[a] Department of Physics, Technical University of Dortmund, 44227 Dortmund, Germany
[b] Department of Physics, University of Augsburg, 86159 Augsburg, Germany
[c] Center for Mathematical Philosophy, Ludwig Maximilian University, 80539 Munich, Germany

[*]Correspondence to: matthias-f.schneider@tu-dortmund.de

**Abstract**

Electric pulses in biological cells (action potentials) have been reported to be accompanied by a propagating cell-surface deformation with a nano-scale amplitude. Typically, this cell surface is covered by external layers of polymer material (extracellular matrix, cell wall material etc.). It was recently demonstrated in excitable plant cells (*Chara Braunii*) that the rigid external layer (cell wall) hinders the underlying deformation. When the cell membrane was separated from the cell wall by osmosis, a mechanical deformation, in the micrometer range, was observed upon excitation of the cell. The underlying mechanism of this mechanical pulse has up to date remained elusive. Herein we report that *Chara* cells can undergo a pearling instability, and when the pearled fragments were excited even larger and more regular cell shape changes were observed (~10—100 $\mu$m in amplitude). These transient cellular deformations were captured by a curvature model that is based on three parameters: surface tension, bending rigidity and pressure difference across the surface. In this paper these parameters are extracted by curve-fitting to the experimental cellular shapes at rest and during excitation. This is a necessary step to identify the mechanical parameters that change during an action potential.

**Keywords:** surface deformation, action potential, pearling, curvature model.

## I. Introduction

Several types of cells, including neurons, myocytes, epithelial as well as some plant cells, propagate all-or-none pulses called *action potentials* (APs) [1–3]. The mechanism that underlies an AP is widely considered to be electrical, and is typically interpreted through a representation of the cell membrane as an equivalent electric circuit [4]. However, evidence of non-electrical aspects that co-propagate with the electric pulse has been accumulating. These include mechanical [5], optical [6] and thermal [7] variations. These results are not explained in the classical framework, and thus have motivated the development of more comprehensive theories [1,8,9]. The focus of this paper is on the mechanical displacement of the surface that co-propagates with the electric pulse. In neurons, mechanical pulses are usually revealed as swelling of the cell cylinder followed by a contraction. These deflections have amplitudes of ~1—10 nm [5]. Several hypotheses were suggested to explain this phenomenon, including: ion and water flow across the membrane [10], distortion of the lipid bilayer due to electrostrictive forces [11], a change in the lipid bilayer state due to a propagating density pulse [8,9] and a volume phase transition in the cortical polymers [1]. Even larger biphasic surface displacements, ~100 nm in amplitude, have been observed during an AP in excitable plant cells (*Characean algae*) [12,13]. These cells are large and easy to handle, and their electrical properties have been widely investigated [14]. Like other plant cells, Characean algae are covered by a rigid cellulose sheath (cell wall). The cell membrane is tightly pressed against this external casing by a large osmotic pressure (turgor) of ~6 bar. By increasing the extracellular osmolarity, the cell membrane can be separated from the cell wall (a process called *plasmolysis*). It has recently been demonstrated that the mechanical pulse that co-propagates with the electric pulse is 10—100 times larger in plasmolysed cells (~1—10 $\mu$m) [15]. These amplitudes are not readily explained by any of the theories that have been previously proposed.

It has been noted previously that plasmolysed *Chara* cells develop a pearling instability [16]. Over time, the cylindrical cell geometry developed a sinusoidal-like shape of dilatations and constrictions. Pearling [17] is an abundant phenomenon that has been observed in elongated cells [18–20] as well as in non-living systems such as lipid membranes [21,22], gels [23,24] and liquids [25,26]. It is typically explained through the Plateau-Rayleigh instability, with the interplay between surface and bulk forces under certain constraints. The most common example is that of a liquid column that collapses into a sphere once its length becomes larger than its circumference. This is a result of the tendency of the liquid to minimize its surface area under a constant volume condition [27]. Lipid membranes, gels and living systems are somewhat more complex, since they have an elastic energy that allows the formation of pearls only above a critical surface tension [21,28]. The dominant parameters in these systems are the surface tension, bending rigidity and transmembrane pressure [29].

As will be shown herein, membrane excitation of pearled *Chara* cells is accompanied by a substantial surface deformation. These deformations are even larger and more regular as compared to plasmolysed cells. Furthermore, the deformations resemble a transient reversal of the pearling process. This provides a good hint that the mechanical changes during an AP may be related to the parameters that govern the pearling condition (surface tension, bending rigidity and transmembrane pressure).

## II. Materials and Methods

*Materials.* All reagents were purchased from Sigma-Aldrich (St. Louis, MO, USA) and were of analytical purity (≥99%).

*Cell cultivation and storage.* *Chara braunii* cells were cultivated in glass aquariums filled with a layer of 2-3 cm of soil, quartz sand and deionized water. The cells were grown under illumination from an aquarium light (14W, Flora Sun Max Plant Growth, Zoo Med Laboratories Inc., San Luis Obispo, CA, USA) at a 14:10 light:dark cycle at room temperature (~20°C). Prior to use, single internodal cells were stored for a minimum of 12 h in a solution containing 0.1 mM NaCl, 0.1 mM KCl and 0.1 mM CaCl$_2$.

*Pearling of Chara internode and excitation of an AP.* A detailed description of the procedure has been published previously [15]. In brief, a single internodal cell (3—6 cm long) was placed on a plexiglass frame into which compartments (~2x5x10 mm) had been milled. The cell was subdivided into electrically isolated sections (length ~5 mm) by perpendicular vacuum grease stripes placed above and below the cell (Dow Corning Corporation, Midland, MI, USA). This allowed for excitation of the cell by means of Ag/AgCl wire electrodes in the compartments. Artificial pond water was added (APW; 1 mM KCl, 1 mM CaCl$_2$, 5 mM HEPES, 110 mM D-sorbitol; pH set to 7.0 with NaOH). After ~10 min, the extracellular osmolarity was gradually increased by addition of APW with higher sorbitol concentrations (initial: ~120 mOsm; final: ~280 mOsm). This led to a gradual efflux of water from the cell (Fig. 1b). When the cell had plasmolyzed (i.e., when the cell membrane had separated from the cell wall), the extracellular solution was replaced by APW (~280 mOsm) containing 50 µM cytochalasin D. This procedure was reported to accelerate the formation of a pearling instability in fibroblasts [19]. Indeed, after 1—2 h, 75% of the cells had pearled (Fig. 1c). Subsequently, the hemispherical region of a pearled cell fragment was monitored by video microscopy (Olympus IX71). Action potentials were triggered by a waveform generator (Agilent 33250A; Agilent, Santa Clara, CA, USA) in combination with a stimulus isolation unit (SIU5; Grass Technologies, Warwick, RI, USA). The membrane potential was monitored by the potassium anesthesia technique [30].

***Extraction of cell surface shapes from video recordings.*** Single frames were extracted from video files. The frame rate of the video recording was ≈10 frames/s, and the field of view was ≈860x650 $\mu m^2$. For each frame, the cellular outline was distinguished from its environment by using a threshold-based segmentation algorithm [31]. The cell surface was subsequently identified by tracing a continuous line of boundary pixels.

## III. Experimental Results
**Surface displacement in pearled cells during APs.** Plant cells are covered by a rigid cellulose sheath, 1—10 $\mu$m thick, called the cell wall. In the native state, an internal pressure (turgor) pushes the cell membrane against the cell wall (Fig. 1a). This pressure difference across the cell surface is due to an osmolarity difference between the intra- and extracellular medium (~270 mOsm vs. ~0 mOsm, respectively). When the extracellular osmolarity was increased gradually by addition of sorbitol, a point was reached when the cell volume started to decrease due to an outflow of water. In this process (plasmolysis), the cell membrane detached from the cell wall (Fig. 1b and Ref. [15]). Within 1—2 hours, the cell formed elongated pearls, within the cell wall cylinder. These pearls were connected by small radii tethers (Figs. 1c and 1d). The radius and length of the pearls were ≈300 $\mu$m and several millimeters respectively. The tether radius was ≈2 $\mu$m. When the cell was electrically excited during pearling formation, the surface motion reversed transiently (Movie S1). These observations suggested that the same parameters that govern the pearling procedure are involved in the mechanical pulse component of an AP.

It was also possible to electrically excite fully formed pearled fragments. Upon stimulation of an AP, a substantial mechanical motion took place on both hemispherical caps of a pearl. The onset of deformation coincided with the depolarization phase of the AP (Fig. 1e), and the time of the peak of deformation varied between experiments, 10—60 s. The deformation of the cell surface was most evident at the connecting region to the tether. Figures 1f and 1g show a close-up of this part of the surface before an electric stimulation and at the peak of the displacement respectively (Movie S2). Interestingly, the pulse did not seem to propagate across the tether into neighboring pearls as evidenced by the absence of mechanical deformation there (Fig. S1 and Movie S3).

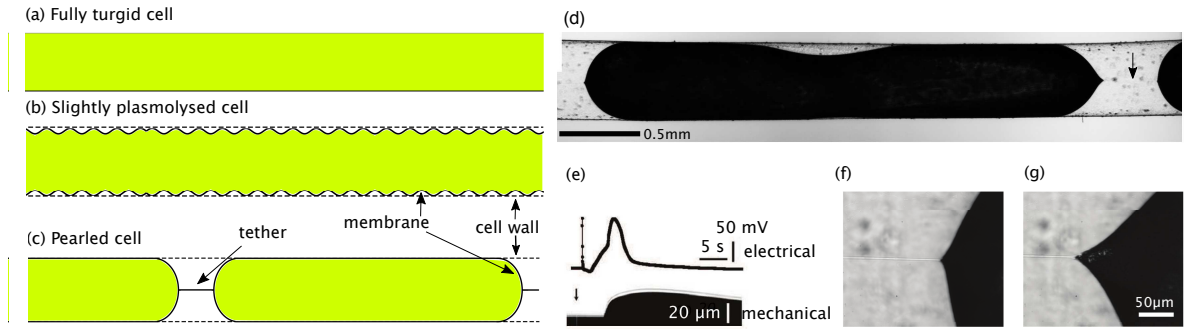

Figure 1: Illustration of (a) fully turgid, (b) slightly plasmolysed, and (c) pearled internodal *Chara* cell. Dashed line represents the cell wall. Solid line represents the cell membrane. (d) Image of pearled *Chara* fragment. Arrow indicates location of tether. (e) Membrane potential pulse (upper) and out of plane displacement of a point on the cell surface close to the tether (lower). Arrow marks time point of stimulation. A close-up of the tether region (f) before stimulation, and (g) at the peak of the surface deformation.

**Extraction of cell surface shapes during an AP.** The shape of a pearl was captured from movies that recorded the entire hemispherical cap during an AP (Movie S4). An example of surface traces for four time points is plotted in Fig. 2a for the cell at rest (t=0) and during various stages of an AP (t=4.9 s, 17.7 s, 54.1 s). The surface underwent a substantial outward displacement with amplitude of ≈30 $\mu$m near the tether (different experiments resulted in varied amplitudes between 10—100 $\mu$m). Conversely, the surface moved inward near the cell wall with an amplitude of ≈5 $\mu$m (inset of Fig. 2a). The outward and inward deformations were separated by a *fixed circle* – a collection of points on the surface of the hemispherical cap that did not move during an AP (see illustration in Fig. 2b). In the subsequent section we use this definition of a fixed circle as a geometrical marker, to investigate partial changes in volume and area during an AP.

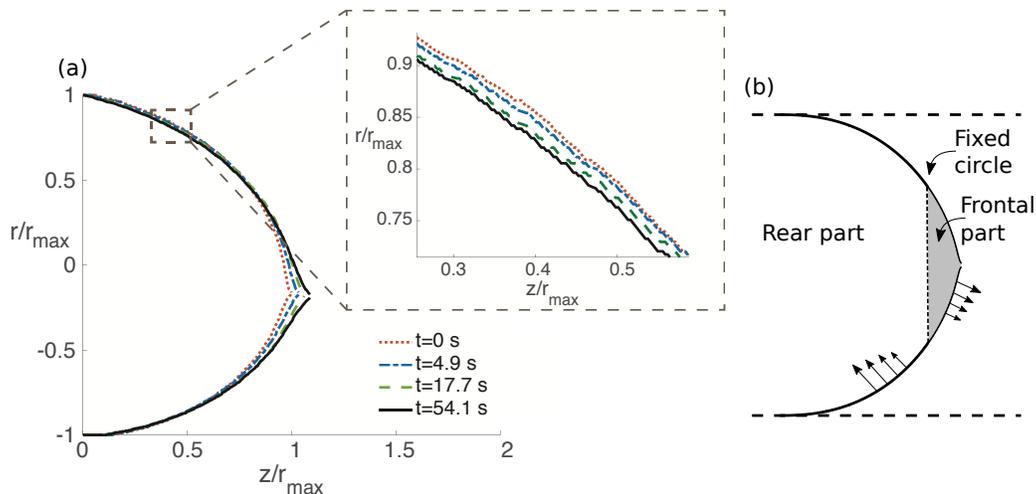

Figure 2: (a) Traces of the cell surface before (t=0) and during various stages of an AP. Inset demonstrates an inward displacement near the cell wall. Cell radius is $r_{max}$=330 $\mu$m. (b) Illustration of the fixed circle (collection of points on the surface that do not move during a deformation) that separates the hemispherical cap into two parts: one which moves outward (frontal) and one which moves inward (rear). Four small arrows indicate the respective direction of deformation.

**Volume and area changes during a surface deformation.** Volume and area of the hemispherical regions of the pearl were calculated by assuming cylindrical symmetry (Eq. (8) below). Furthermore, these were also calculated separately for the two parts distinguished by the fixed circle (Fig. 2b). The results are plotted in Fig. 3 as relative values to the resting state (t=0). The total change in volume and area of the hemispherical cap was less than 2%. The volume of the frontal part of the pearl increased by 21% and its area increased by 8%. Apparently this was compensated by a decrease in volume and area of the rear part. These results repeated themselves in trend and order of magnitude in n=4 cells. These observations indicate that the shape changes are not simply due to a decrease or increase of the pearl volume, but rather due to a shift of intracellular volume.

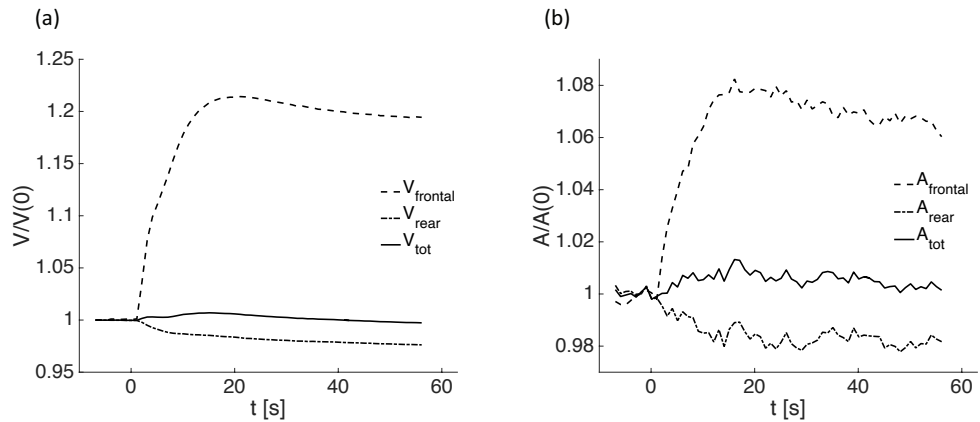

Figure 3: Total (a) volume and (b) area of the hemispherical cap as a function of time (solid line), upon excitation of an AP at t=0. Values were normalized by the resting state value (t=0). Volume and area were also calculated separately for the frontal (dashed line) and rear (dashed-dotted line) parts of the hemispherical cap (see Fig. 2b).

## IV. Theoretical Model and Analysis

**Model description.** When a *Chara* cell is excited, a membrane potential pulse is generated and propagates along the cell cylinder. In parallel, a cell surface deformation pulse occurs. The mechanical and the electrical signals propagate together along a *Chara* cell at a velocity of ~1 cm/s (Fig. 1e and Ref. [15]). In the present work we do not attempt to provide detailed dynamical equations that describe the propagation of the mechanical pulse. Rather, our objective is to understand its underlying mechanism. We hypothesize that the deformation of the surface may be explained by a reversible change in its mechanical properties.

The system under consideration is the cellular surface, represented as a two-dimensional contour embedded in a three-dimensional space. We consider three macroscopic parameters to describe its mechanical properties: the surface tension $\sigma$, the bending rigidity $\kappa$ and the pressure difference across the surface $\Delta p \equiv p_{out} - p_{in}$ [29]. Clearly this is an over-simplification since the surface is not an infinitesimally thin fluid. Rather, it consists of the cell membrane, cortical cytoskeleton, chloroplasts and subcortical actin bundles [32]. Nonetheless, our results demonstrate that this minimum set of parameters captures the shape transformations in a very satisfactory manner. We return to this issue in the Discussion section. Under these assumptions, the elastic energy of the surface is given by

$$E = \int \sigma dA + \int \kappa dH + \int \Delta p dV, \quad (1)$$

with $A$ the surface area, $H$ the mean curvature of the surface, and $V$ the enclosed volume [29]. A Gaussian curvature term was dropped from the equation, since it contributes a deformation-independent term.

We treat the mechanical parameters as constants in space, but not in time. This assumption is valid when the pulse length is much larger than the field of observation. The spatial extension of an AP can be estimated from the pulse velocity (~1 cm/s) and duration (~10 s) to be ~10 cm. Indeed, this is significantly larger than the field of observation (~1 mm). In addition, only the linear regime of the bending modulus is considered. Under these additional assumptions, the energy function becomes

$$E = \sigma \int dA + \frac{\kappa}{2} \int (2H)^2 dA + \Delta p \int dV. \quad (2)$$

At the cell wall additional forces are induced by the rigid structure (Fig. 1c), and at the tether the surface is likely balanced by intracellular polymers (see Discussion section and Ref. [33]). Thus, the model is valid only in the hemispherical cap region, and the connections with the cell wall and tether are treated as boundary conditions (Eq. (7) below).

In order to estimate the mechanical change of the surface, we make a simplified assumption that the rate of change of the parameters is slower than the surface response [34]. This implies that at each time frame the curves are extremum solutions of the energy function. We discuss the validity of this assumption in the Discussion section. The Euler-Lagrange equation that minimizes the above energy function was calculated by Zhong-Can and Helfrich [35]

$$\Delta p - 2\sigma H + 2\kappa(2H^3 + \nabla^2 H) = 0. \quad (3)$$

To keep calculations simple, only axially symmetric solutions of the above equation are considered.

Parametrization of the surface curve is shown in Fig. 4. The internal coordinate of the surface is represented by the arclength $s$ along the contour. The external cylindrical coordinates $(r, z)$ are related to the arclength by [36]

$$\begin{aligned} r_s &= \cos(\psi), \\ z_s &= -\sin(\psi), \end{aligned} \quad (4)$$

where $r_s \equiv \frac{dr}{ds}$ and $\psi$ is the angle between the tangent to the contour and the $r$ axis.

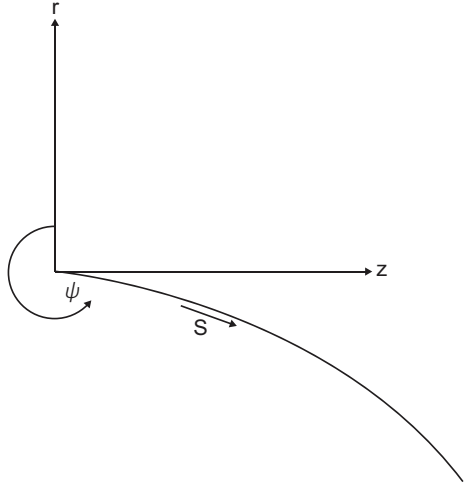

Figure 4: Parametrization of the surface curve.

The curvature of axisymmetric curves satisfies [37]

$$H = \frac{1}{2}\left(\frac{r_{ss}}{z_s} - \frac{z_s}{r}\right) = \frac{1}{2}\left(\psi_s + \frac{\sin(\psi)}{r}\right), \quad (5)$$

and the Laplacian operator is

$$\nabla^2 H = \frac{1}{r}\frac{d}{ds}(rH_s). \quad (6)$$

Four boundary conditions describe the connection point with the cell wall and the tether:

$$r(0) = r_{max}, \quad \frac{dr(0)}{dz} = 0,$$
$$r(\text{end}) = r_{min}, \quad \frac{dr(\text{end})}{dz} = 0, \quad (7)$$

where $r_{max}$ is the radius of the cell wall and $r_{min}$ is the radius of the tether. These four conditions determine a unique solution that depends on three dimensionless parameters: $h \equiv \frac{\Delta p r_{max}}{2\sigma}$ (pressure-to-tension parameter), $\epsilon \equiv \frac{\kappa}{\sigma r_{max}^2}$ (bending-to-tension parameter) and $\frac{S}{r_{max}}$ (curve length). Thus, the space of solutions is three-dimensional. The Zhong-Can—Helfrich model (Eqs. (3)—(7)) was solved using standard relaxation methods [38].

Other geometrical quantities of use are the volume $V$, area $A$, and axial length $L$

$$V = -\int \pi r^2 \sin(\psi)\, ds,$$
$$A = \int 2\pi r\, ds, \quad (8)$$
$$L = -\int \sin(\psi)\, ds.$$

Finally, curve solutions were fitted to experimental traces and were evaluated by a least squares error function

$$LS_{err} = \frac{1}{L} \int \Delta r^2 dz \quad (9)$$

where $\Delta r$ is the difference between the experimental trace and the calculated curve.

**Comparison of surface solutions to experimental results.** In order to calculate surface traces, we have used an idealized curvature model which is governed by three mechanical parameters: surface tension $\sigma$, bending rigidity $\kappa$ and pressure difference across the surface $\Delta p$. Before describing its results, one may ask if a simpler model may be sufficient to capture the surface deformation. It was previously suggested that a curvature model with negligible rigidity ($\kappa = 0$) may be used to model stretched-induced pearling in axons [39]. Thus, solutions were first explored for this simplified version which reduces the equilibrium equation into the Young—Laplace equation [40]. Indeed, solutions roughly approximated the pearling formation process (Fig. S2). However, these solutions did not describe fully developed pearls connected by tethers with small radii (see also Ref. [37]). Therefore, to capture the experimental traces it was necessary to solve the full Zhong-Can—Helfrich model (Eqs. (3)—(7)); i.e., an elastic energy equation that includes a non-zero bending rigidity.

The Zhong-Can—Helfrich model depends on three dimensionless parameters: $h \equiv \frac{\Delta p r_{max}}{2\sigma}$ (pressure-to-tension parameter), $\epsilon \equiv \frac{\kappa}{\sigma r_{max}^2}$ (bending-to-tension parameter) and $\frac{S}{r_{max}}$ (curve length). For a given curve length, solutions were found in a limited region of the $(h, \epsilon)$ phase space. A comparison between experimental traces of the surface and the calculated curves is plotted in Fig 5a. Two traces, before stimulation (solid black line) and at the peak of the deformation (dashed black line), are overlaid with solutions of the Zhong-Can—Helfrich model (solid and dashed grey lines respectively). This constitutes evidence that the mechanical parameters ($\Delta p, \sigma$ and $\kappa$) are in principle sufficient to provide a satisfactory description of the shape of the surface.

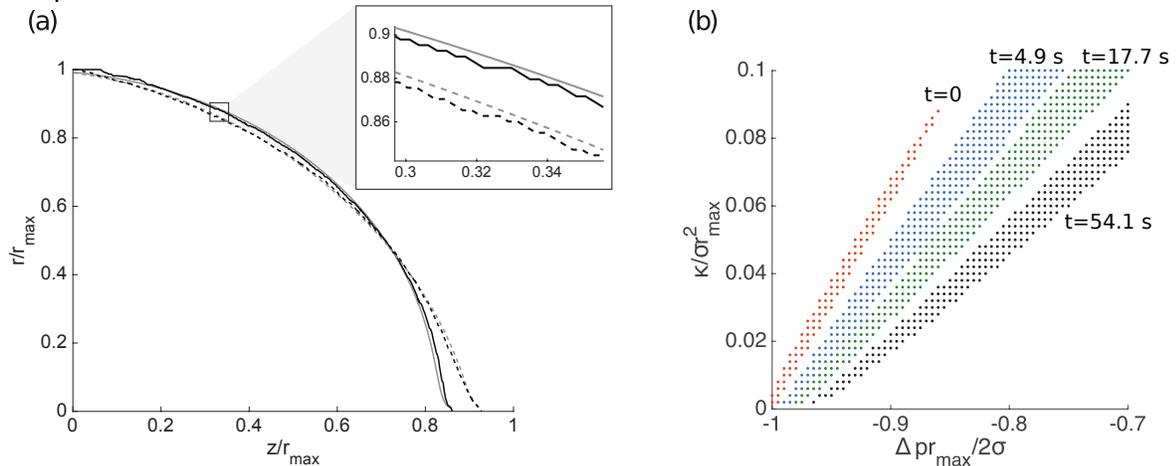

Figure 5: (a) Experimental traces of the hemispherical cap, before stimulation (solid black line) and at the peak of the deformation during an AP (dashed black line). These are compared with solutions of the Zhong-Can—Helfrich equation (solid and dashed grey lines, respectively). Parameters of the curved solutions are given in Ref. [40]. (b) Multiple parameter combinations which led to a good fit of the experimental shapes are plotted in phase space. The four areas correspond to successful fits to the four contours shown in Fig. 2a (error function $<1.5 \cdot 10^{-4}$, Eq. (9)).

The experimental traces are well captured in a restricted region of the $(h, \epsilon)$ phase space (Fig. 5b). Each point in the two-dimensional phase space represents a state of the surface and is related to a particular shape. It is impossible to relate an experimental trace to a unique state, because different pairs of $(h, \epsilon)$ yield very similar curves. This suggests a compensation effect between $h$ and $\epsilon$ in this regime of phase space. To estimate values of the physical parameters (surface tension, bending rigidity and pressure difference across the surface) we use a previously measured value of the surface tension of a plasmolysed *Chara* cell membrane, $\sigma \sim 10^{-5} \frac{N}{m}$, and the cell wall radius, $r_{max} \sim 10^{-4}\ m$ [15]. In the regime of well-fitted solutions in phase space (h~-1 and $\epsilon$ ~0.01—0.1) this corresponded to $\Delta p \sim 0.1\ N/m^2$ and $\kappa \sim 10^{-15} - 10^{-14}\ J$. In the Discussion section we clarify the relatively large value of the bending rigidity, as compared to a lipid bilayer membrane ($\kappa_{lbm} \sim 10^{-19}\ J$).

## V. Discussion

Surface displacements during an AP were typically considered to be of nano-scale order [5]. However, it was recently demonstrated in *Chara* cells that by separating the cell surface from the cell wall, the actual surface deformation is 10—100 times larger in amplitude [15]. In this paper we continue this study and report that even larger and more regular deformations are observed in fully developed pearls of *Chara* cells.

### Shapes of pearled excitable plant cells from elastic energy models.

Experimental traces of cellular shape during excitation were analyzed in the frame of the Young—Laplace and the Zhong-Can—Helfrich models. The absence of bending rigidity in the Young—Laplace model did not allow the formation of small tethers. This has been noted by others previously [37]. On the other hand, satisfactory solutions that captured the cell surface traces were obtained in the Zhong-Can—Helfrich model. This indicates that the state of the surface can be represented by three surface properties ($\Delta p, \sigma, \kappa$). Although the Zhong-Can—Helfrich model provides a satisfactory description of the cell surface during cellular excitation, it was not possible to identify a unique trajectory in the mechanical phase space. This is apparently due to a compensation effect between the model parameters ($h \equiv \frac{\Delta p r_{max}}{2\sigma}, \epsilon \equiv \frac{\kappa}{\sigma r_{max}^2}$). In order to capture the surface deformation during an AP, one can, for instance, decrease $\epsilon$ fivefold at constant $h$ (Fig. S3a). This would occur, for example, if the bending rigidity decreased fivefold at constant surface tension and pressure difference. A very similar shape change is induced by decreasing the absolute value of $h$ by 15% at constant $\epsilon$ (Fig. S3b). This would occur, for example if the pressure difference decreased. Clearly, other trajectories are possible by combining changes in $\sigma, \kappa$ and $\Delta p$ (Fig. 5b).

In order to obtain valid solutions of the surface, larger bending rigidities, as compared to typical lipid bilayer membranes, were required ($\kappa \sim (10^4 - 10^5)\kappa_{lbm}$). While our analysis deals with an infinitesimally thin surface, the cellular cortex of a *Chara* internode consists not only of a plasma membrane, but also of a cortical cytoskeleton, chloroplasts and subcortical actin bundles [32]. Thus, the higher rigidity found in our analysis simply reflects a higher overall bending stiffness of the composite surface. To demonstrate that, let us recall that the bending rigidity is related to the Young's modulus according to [41]

$$\kappa \sim \frac{1}{10} Y d^3 \quad (10)$$

with $Y$ the Young's modulus and $d$ the thickness of the layer. For a lipid bilayer membrane, $d \sim 5$ nm and $Y \sim 10^7 \, Pa$ [41]. The estimated bending rigidity is therefore $\kappa_{lbm} \sim 10^{-19} \, J$, which agrees well with measured values [42]. For the cell cortex, the measured Young's modulus is rather smaller than that of a lipid bilayer, $Y \sim 10^3 \, Pa$ [43]. By considering a surface thickness of $d \sim 5 \, \mu$m, the estimated bending rigidity is $\kappa \sim 10^{-14} \, J$, which agrees with our curve-fitting results ($10^{-15} - 10^{-14}$ J). It should be noted that additional elastic parameters, that were not considered in this simple model, may also change during an AP (for example the in-plane shear rigidity and the surface compressibility). Therefore, it may be worth to investigate a more elaborate model of the cell surface elasticity in a future work.

Another point of interest was that the tether in pearled *Chara* protoplasts (~1 μm in radius) was larger compared to other cell membrane tethers, for instance in erythrocytes (~100 nm) [44]. One potential reason for this larger diameter may be the inclusion of cytoskeletal filaments. It can be estimated that $\sim 10^5$ actin filaments, with a diameter of $\sim 5$ nm, run along the axial direction of a *Chara* cell [45]. If these filaments are enclosed within the tether, this will limit its radius to $\gtrsim 1 \, \mu m$, similar to observations (Fig. 1f). In pearled neurons it was indeed shown that the tether region contains a high number of filaments [33]. An indication of the existence of filaments inside the tether of a pearled *Chara* cell is shown in Movie S5. In this experiment, application of voltage led to rupture of the cell membrane and disintegration of the pearl. In the tether region, there is an apparent recoiling of small filamentous structures, which are most likely cytoskeletal filaments.

**Dynamics of shape transformation during cellular excitation.**
One of the main assumptions of the model is that the response of the surface is faster than the change of the surface parameters. Let us now estimate the characteristic response time of the surface by comparing the elastic forces with the viscous resistance of the bulk [34]. For inflating, stretching or bending a surface it is $\tau_{\Delta p} \sim \frac{\eta}{\Delta p}$, $\tau_\sigma \sim \frac{\eta r_{max}}{\sigma}$ and $\tau_\kappa \sim \frac{\eta r_{max}^3}{\kappa}$, respectively. In a *Chara* cell, the intracellular bulk is composed of a large vacuole filled with an aqueous solution of ions and low molecular weight substances (cell sap). Assuming that the cell sap has a viscosity of water $\sim 10^{-3} \, Pa \cdot s$, the response time scales are estimated as $\tau_{\Delta p} \sim \tau_\sigma \sim 10^{-2} \, s$ and $\tau_\kappa \sim 1 \, s$. In comparison, the duration of an AP in *Chara* cells is ≈5—10 s. Thus, the use of a quasi-static analysis is not unreasonable. This is not to say that the response of the surface is negligible. Neither the viscous properties of the plasma membrane nor of the cortical layers are properly reflected in this simple estimate. It is reasonable to expect that these properties increase the response time of the surface. In this regard, an interesting point for future research will be to clarify why the deformation outlasts the voltage pulse in time (Fig. 1e). One possibility is that the change in a mechanical parameter of the surface is proportional to the change in membrane potential. If this is the case, a delay in the surface response could, for example, be due to a viscous effect. However, it may also be the case that the voltage and the mechanical parameters are coupled but have different time scales. Therefore, it will be very beneficial to study the viscoelastic response of the cell surface in detail.

**Theories of the AP in relation to mechanical state changes.**
We now turn to examine how mechanical changes of the cell surface are treated within different theories of the action potential. The classical theory of cellular excitation (the Hodgkin and Huxley (HH) model) does not directly deal with a mechanical aspect [4]. However, a main mechanism of the HH model is transmembrane flux of ions, and some have attempted to interpret mechanical deformations in neurons solely based on volume changes induced by osmosis [10]. Indeed, excitation in *Chara* is associated with an influx of $Ca^{2+}$ and an efflux of $Cl^-$ and $K^+$ ions add citation, e.g. Oda 1976; Ashley Williamson 1982. In addition, it was reported that a slight decrease in volume occurs during an AP [46], from which it can be estimated that the change of volume is $\frac{\Delta V}{V} \sim 10^{-4}$. Therefore, this common view point suggests the following cause and effect chain. (i) Transmembrane flux of ions during an AP induces an osmotic pressure difference across the membrane. (ii) The pressure gradient is subsequently compensated by an outflow of water. (iii) Water outflow reduces the cell volume. (iv) Due to volume reduction, the surface of a *Chara* cell is predicted to move inward. Since the transmembrane pressure is rapidly compensated by a flux of water, it is *itself* not involved as a force that drives the displacement. In contrast, our measurements in *Chara* cells have showed that the surface deforms both inward and outward (Fig. 2 and Ref. [15]). We have further demonstrated that the intracellular advection is significantly larger than the previously measured volume loss (Fig. 3). These observations cannot be explained without considering the mechanical forces that act on the surface. In conclusion, it is conceivable that changes in transmembrane pressure drive the surface deformation, by virtue of its direct action on the surface interface. However, in order for that to occur, water flow across the membrane *must not* cancel it, in contrast to the common view point.

More comprehensive theories of cellular excitation treat the pulse as a propagating thermodynamic state change [1,8,9]. The second law of thermodynamics provides a natural coupling between observables, in particular, an electro-mechanical coupling during pulse propagation [47]. Indeed, density pulses in lipid bilayer membranes are associated with electrical changes, as well as changes of surface tension and bending rigidity [47,48]. The present work has shown that changes in these elastic parameters can lead to shape changes as observed in *Chara* pearls. Ideally, in a next step one will measure $\sigma, \Delta p$ and/or $\kappa$ during an AP. This will allow to identify the trajectory in phase space that the system takes during cellular excitation. Additionally, it will be crucial in future work to establish state diagrams of excitable cell surfaces. This will reveal if one of the fundamental requirements for a non-linear density pulse; i.e., a non-linearity in the thermodynamic susceptibilities of the system [49], is present in excitable cell surfaces. Finally, it will be important to address if the submembranous filaments in excitable cells play an active role in the surface deformation, as was proposed previously [50].

**Acknowledgements**

The authors thank Konrad Kaufmann, Jan Kierfeld and Jonas Hegemann for fruitful discussions, Asaf Liberman for graphical assistance, and the two anonymous reviewers for their suggestions and comments. MFS acknowledges financial support by the German Science Foundation (DFG) as well as the research unit SHENC.